\documentclass[12pt]{article}

\begin{document}

\vspace*{-.6in}
\thispagestyle{empty}
\begin{flushright}
DAMTP-1998-124\\
hep-th/9809043
\end{flushright}
\baselineskip = 20pt

\vspace{.5in}
{\Large\bfseries
\begin{center}
Compactification of Supermembranes
\end{center}}
\vspace{.4in}

\begin{center}
C. Codirla\footnote{Email: cc10011@damtp.cam.ac.uk.} and
M. J. Perry\footnote{Email: Malcolm@damtp.cam.ac.uk.}\\
\emph{D.A.M.T.P., University of Cambridge, Cambridge CB3 9EW, U.K.}\\
\today
\end{center}
\vspace{.3in}

\begin{center}
\textbf{Abstract}
\end{center}
\begin{quotation}
  We look at the vertical dimensional reduction of the supermembrane of 
  M-theory to the D2-brane of Type IIA string theory.
  Our approach considers the soliton solutions of the two low energy field
  limits, $D=11$ and $D=10$ Type IIA supergravities, rather than the 
  worldvolume actions.
  It is thus necessary to create 
  a periodic array.  The standard Kaluza-Klein procedure requires that the 
  brane is smeared over a transverse direction, but we will keep the 
  dependence on the compactification coordinate, seeing how the eleventh 
  dimension comes into play when we close up on the D2-brane.

\end{quotation}
\vfil
\newpage
  
\section{Introduction} \label{sec:introduction}  
  String theory has undergone a dramatic change from five separate theories,
  to the realization that they all are limit points in the moduli space
  of an underlying fundamental theory, dubbed M-theory. 
  The five string theories are non-perturbatively equivalent, and related through various
  duality symmetries. M-theory is 
  intrinsically eleven-dimensional, with $D=11, N=1$ supergravity as its
  low energy limit.  It is also the strong coupling limit of Type IIA string
  theory\cite{Witten:95}.
  Since we are unable to perform quantum mechanics in the eleven dimensional 
  vacuum it is necessary to look at the BPS spectrum of 
  M-theory to test and derive dualities between the various 
  string theories in different dimensions.  We can rely on such manipulations,
  since these states receive no quantum corrections to their masses, as long
  as supersymmetry remains unbroken \cite{WitOli:78}.

  The fundamental BPS-states of M-theory are the supermembrane 
  \cite{BerSezTow:87,DufSte:91} and the superfivebrane \cite{Guven:92}, both
  of which appear as solutions to M-theory's low energy approximation, 
  $D=11, N=1$ supergravity.  Relying on the interconnectedness of all 
  theories the speculation is put forward in \cite{Townsend:95} that these
  solutions are directly related to the extended objects of Type IIA string
  theory \cite{DabGibHarRui:90,DufLu:92,DufLu:94,DufKhuLu:94,Stelle:97}.
  The relation to the other string theories is established via dimensional
  reduction and U-dualities.  All these extended objects, being BPS states 
  also appear as solutions of the low energy field theories of these string 
  theories.

  The aim of this paper is to explicitly look at the relation between the
  supermembrane\footnote{From now on referred to as the M2-brane} 
  and the D2-brane.
  There exist two approaches; the first concentrates on the world-volume
  action \cite{Townsend:95b,DufLu:92}.  The world-volume is three-dimensional
  giving a duality between a scalar, usually the eleventh coordinate, and a vector
  $A_{\mu}$, the world-volume gauge-field of the resulting D2-brane.  In 
  this way the eleventh dimension is eliminated via a Lagrange multiplier, 
  $A_{\mu}$. It is surprising that the classical supermembrane reproduces 
  the D2-brane action, which comes from a one-loop sigma model calculation 
  in string theory \cite{Townsend:95b,Leigh:89}.

  The second approach looks at low energy representations of the BPS states
  as solutions to supergravity theories.  These can be obtained mostly from
  the $D=11$ version through standard Kaluza-Klein reduction.  As this
  reduction is consistent the lower dimensional solutions also satisfy
  the higher dimensional field equations.  However it is also possible to
  reduce the solutions directly \cite{LuPopSezSte:95}, in two different ways.
  The more common one takes a $(D,p)$\footnote{$D$ is the spacetime 
  dimension, and $p$ the spatial extent of the $p$-brane} solution to 
  $(D-1,p-1)$ solution, by simultaneously reducing a spatial and 
  world-volume direction.  This follows closely the procedure also 
  applicable to the world-volume approach \cite{DufHowInaSte:87,DufSte:91}.
  The second, more intricate reduction, takes a $(D,p)$ to a $(D,p-1)$ 
  solution, which is what we will be looking at here.  Such
  reductions have been discussed previously in Refs  
  \cite{Myers:87,GauHar:92,Khuri:95,Welch:95,LuPopSte:96,LuPopXu:96,LuPop:97b,
        Stelle:97}.
        
  Looking at the
  movement of the solutions on a $(D,p)$ plot, we see that the former method
  moves the solution diagonally down, whereas the latter has a vertical
  movement.  Hence the two procedures are called diagonal and vertical 
  reduction respectively.  E.g.~diagonal reduction takes the M5-brane to
  the D4-brane of IIA string theory, and vertical to the NS5-brane.

  Diagonal Kaluza-Klein reduction is more readily performed due to the 
  independence of the supergravity fields on the world-volume coordinates.
  Such an isometry must first be created in the transverse direction for
  vertical reduction.  This method is referred to as ``construction of 
  periodic  arrays''.  Using the no force condition between BPS states, we
  create an infinite array in a direction transverse to our $p$-brane.
  The solution can then be viewed as a compactification along this particular
  direction, say ${\bf{\hat z}}$.  The dependence on $z$ however poses a 
  problem for the orthodox Kaluza-Klein procedure.  Therefore the original
  construction is modified by letting the distance between the original
  branes go to zero.  This smears out the brane over ${\bf{\hat
  z}}$, removing the dependence on the transverse coordinate.
  Performing the reduction and conformally rescaling the metric to go 
  back to the Einstein frame of $(D-1)$ supergravity gives the
  $(D-1,p)$ brane. When oxidizing the $(D-1,p)$ brane to $D$ dimensions the
  solution can be naively interpreted as the above smeared brane \cite{LuPop:96b}.
  However, from the string point of view, it makes more sense to reinterpret 
  the oxidation as a periodic array of string solitons, which is certainly 
  another possibility of the oxidation \cite{LuPop:97b}.

  The limiting procedure of smearing out the brane along a transverse
  direction hides the fact that we are dropping the heavy modes in this
  construction which arise from the compactification along $\hat {\bf z}$. 
  It is these that die off exponentially as the radius is shrunk,
  playing a role in eliminating the singularity structure
  encountered in most of the IIA supergravity solutions.  
  We will see this happen explicitly for the case at hand M2 
  $\rightarrow$ D2-brane.

  The paper is organized as follows: section \ref{sec:pbranerev} reviews the 
  construction of $p$-branes in $D=11$ and $D=10$ Type IIA supergravity
  theories and their dimensional reduction via the orthodox Kaluza-Klein
  method.  Next we explicitly construct the stacked M2-brane and use the periodic
  array construction to obtain the D2-brane, comparing the discretized 
  to the smeared brane reduction.  In section \ref{sec:10dptofv}
  we discuss the ten dimensional point of view of the reduction, resolving 
  the eleventh dimension at the horizon of the D2-brane in the next section,
  where we also look at the global eleven dimensional spacetime, 
  ${\bf R}^{10} \times S^1$.

\section{$p$-branes revisited} \label{sec:pbranerev}
 The $p$-brane solutions we will be interested in are solutions to supergravity
 theories derivable from $D=11, N=1$ supergravity.  We therefore start off
 from here to find the M2-brane. Next we Kaluza-Klein reduce the Lagrangian
 to obtain the extremal solutions of IIA supergravity, the BPS states of 
 IIA string theory.
 
 All brane solutions we will be considering are bosonic in nature.  Therefore
 we can ignore the fermionic sector, needing only the fermionic 
 transformation laws to ensure that the solutions are indeed supersymmetric.
 The $D=11$ supermultiplet contains the graviton described by $e_M{}^A$,
 the gravitino, a Rarita-Schwinger vector-spinor, $\Psi_M$, and the 3-form 
 gauge field $A_{MNP}$, with field strength 
 $G_{MNPQ}=4\partial_{[M}A_{NPQ]}$.  The bosonic Lagrangian resulting is
 \cite{CreJulSch:78}
\begin{equation} \label{11sugra}
 \mathcal{L}  = \sqrt{-g} \left[ R - \frac{1}{48} G^{MNPQ}G_{MNPQ}\right]
    + \frac{1}{6\cdot 3!4!^2} \epsilon^{M_{1} \ldots M_{11} } G_{M_{1}\ldots
    M_{4}} G_{M_{5}\ldots M_{8}} A_{M_9 \ldots M_{11}}. \nonumber
\end{equation}
  Note that the theory has no dilaton and does not suffer from
  ambiguities of rescaling the metric by a conformal factor.  Further, the
  solutions we will consider make the variation of the Chern-Simons term vanish, 
  so that we can ignore it from now on.

  One can now perform Kaluza-Klein reduction to type IIA supergravity using 
  the ansatz for the metric
\begin{equation} \label{reductionmetric}
  ds_{11}^2=e^{2\alpha\varphi}ds_{10}^2+e^{2\beta\varphi}(dz+
            \mathcal{A}_M dx^M)^2,
\end{equation}
  where the dilaton $\varphi$, the Kaluza-Klein vector field 
  $\mathcal{A}$, and the ten-dimensional metric $ds_{10}^2$ are taken to be 
  independent of the extra coordinate $z$.  
  The gauge field $A_{MNP}$ is split up into a 2-form and 3-form, both also 
  independent of $z$.  The two constants $\alpha, \beta$ are chosen so that the 
  resulting lower dimensional theory is in the Einstein frame:
  $\alpha=-\frac{1}{12},\beta=\frac{2}{3}$. The resulting Lagrangian has 
  the same field content as IIA supergravity\cite{LuPopSezSte:95}.

  One can now proceed to look for extended brane solutions to both these
  theories.  The ansatz for the metric and potential in either 
  $D=10$ or $D=11$ is given by \cite{DabGibHarRui:90,DufSte:91}
\begin{eqnarray} 
  ds_D^2&=&e^{2A}dx^{\mu}dx^{\nu}\eta_{\mu\nu}+e^{2B}dy^m dy^n\delta_{mn}\label{braneansatz}\\
  A_{0 1 \ldots (d-1)}&=&e^C,\label{Potansatz}
\end{eqnarray}
 where $x^{\mu}{}(\mu=0\ldots d-1)$ are the world-volume coordinates of the
 $(d-1)$-brane, and $y^m{}(m=1\ldots (D-d))$ are the transverse space 
 coordinates.  The functions $A, B, C$ are all functions of the transverse 
 radial distance $r=\sqrt{y^m y^m}$.  In $D=10$ we write the dilaton as
\begin{equation}
 \phi=\phi(r).
\end{equation}

  One now imposes supersymmetry on these solutions by looking at the 
  transformation of the fermionic sector of the theory.  For the eleven-
  dimensional theory we have $\delta\Psi_M=\tilde{D}_M\epsilon=0$:
\begin{equation} \label{sugratrans}                                                              
\tilde{D}_M\epsilon=\left(\partial_M+\frac{1}{4}\omega_M{}^{AB}\Gamma_{AB}  -
\frac{1}{288}(\Gamma^{PQRS}{}_M+8\Gamma^{PQR}\delta^S{}_M)G_{PQRS}\right)\epsilon=0
\end{equation}                                                                
 \cite{CreJulSch:78,DufSte:91}.  In (\ref{sugratrans}) $\omega_M{}^{AB}$ is 
 the spin connection and the Dirac matrices are defined as
 $\Gamma^{A_1\ldots A_N}=\Gamma^{[A_1}\ldots\Gamma^{A_N]}$, antisymmetrized
 with weight one, where $M,N$ are spacetime indices and $A,B$ are tangent
 space indices.
  
  Same laws exist for the type IIA supergravity theory.  To find such
 an $\epsilon$ one splits up this spinor according to the
  global symmetry (Poincare)${}_d$ $\times$ SO($D-d$), 
  $\epsilon(y)=\epsilon\otimes\eta(y)$; and similarly for
  the $\Gamma$ matrices.  Inserting the ans\"{a}tze (\ref{braneansatz}),
  (\ref{Potansatz}) into the transformation laws
  gives us $A(r)$ and $B(r)$ linearly dependent on $C(r)$. Insertion
 into the equations of motion gives:  
\begin{equation} \label{laplace}
  \delta^{mn}\partial_{m}\partial_{n}e^{-C}=0
\end{equation}
  \cite{DabGibHarRui:90,DufSte:91,DufKhuLu:94}. 
  Laplace's equation is solved by
\begin{equation} \label{solution}
  e^{-C}=1+\frac{k_d}{r^{D-d-2}},r>0.
\end{equation}
  The ansatz also reduces 
  $\eta(r)$ to $f(e^{-C})\eta_0$, $\eta_0$ a constant spinor, which needs to be
  projected into a chiral eigenstate.  The number of supersymmetries is
  halved, leaving us with a supersymmetric $p$-brane.  
  
  The final solution to the two theories considered above is now given by
\begin{eqnarray} \label{pbranesoln}
  ds^2&=&(1+\frac{k_d}{r^{\tilde{d}}})^{-\frac{4\tilde{d}}{\Delta(D-2)}}               
          dx^{\mu}dx^{\nu}\eta_{\mu\nu} +
       (1+\frac{k_d}{r^{\tilde{d}}})^{\frac{4 d}{\Delta(D-2)}}
          dy^m dy^m \nonumber \\
  e^{\phi}&=&(1+\frac{k_d}{r^{\tilde{d}}})^{\frac{2 a}{\Delta}},
\end{eqnarray}
  where $\tilde{d}=D-d-2$ and $a^2=\Delta-\frac{2d\tilde{d}}{D-2}$, $\Delta=4$
  in $D=10$ and $D=11$ \cite{LuPopSezSte:95}. For $D=11$ 
  we see that $a=0$, which reflects the absence of any dilaton 
  fields in this dimension.

  The supersymmetry of these solutions implies the saturation of a 
  Bogomol'nyi bound \cite{DabGibHarRui:90,DufSte:91}, which is established 
  by evaluating both the Noether ``electric'' charge $\mathcal{Q}$ and the 
  mass per unit volume, $\mathcal{M}$, obtaining
\begin{equation}
  \mathcal{Q}=\frac{1}{4\Omega_{\tilde{d}+1}}\int_{ {\bf S}^{\tilde{d}+1}}
  (*G_n)\ge \mathcal{M}.
\end{equation}
  Here $\Omega_{\tilde{d}+1}$ is the volume of the sphere living at 
  the boundary of the transverse space of the $p$-brane under consideration.  
  To calculate the ADM mass we look at the first order 
  perturbation of the metric $g_{MN}$, 
  $h_{MN}=g_{MN}-\eta_{MN}$, which falls off like $O\left(\frac{1}{r}\right)$.  
  Using cartesian coordinates, and letting $a=1\ldots D-1$ and $m$ run over 
  transverse coordinates, the form of $\mathcal{M}$ is given by
\begin{equation}  \label{ADMmass}                     
  \mathcal{M}=\frac{1}{4\Omega_{\tilde{d}+1}}\int_{ {\bf S}^{\tilde{d}+1}}
  d^{\tilde{d}+1}\Sigma^m(\partial^n h_{mn} - \partial_m h^a{}_a)
\end{equation}
  \cite{MisThoWhe:73,Lu:93,Stelle:97}.

  Furthermore the mass can be related to the integration constant $k_d$ by
  noticing that Eq.~(\ref{laplace}) is not exactly satisfied by 
  Eq.~(\ref{solution}); it produces a delta function.  This hints at the 
  possibility of a source at $r=0$, which for a $p$-brane can be coupled to the 
  supergravity action, $I_D$ via $I_d$, where
\begin{equation}  \label{coupledbrane}
  I_d= -T_d\int d^d\xi \sqrt{-\gamma}-T_d\int_{world-volume}A_d.
\end{equation}
  The equation of motion (\ref{laplace}) now reads
\begin{equation}
  \delta^{mn}\partial_m\partial_n e^{-C}=2T_d\delta^{D-d}(y),
\end{equation}
  where $d=p+1$ is the worldvolume dimension and $\gamma$ is the 
  worldvolume metric,
  giving $k_d=2 T_d/\tilde{d}\Omega_{\tilde{d}+1}$.\footnote{
  Throughout we have set Newton's constant of gravity, $\kappa^2_{11}=1$}

\section{Vertical Dimensional Reduction of M2-brane} \label{m2reduc}
  Having given the form of the solutions for the M2 and D2 brane in $D=11$ 
  and $D=10$ dimensions respectively, we construct the periodic array
  necessary for the vertical dimensional reduction.  This will give
  us an isometry in a transverse direction, which can then be eliminated
  via the standard Kaluza-Klein procedure.

  To construct such an array along a chosen transverse axis, the $p$-brane
  solutions must remain static, i.e.~fixed in their location.  This is 
  possible due to the no-force condition, which gives us zero interaction
  between the soliton solutions.  This can be verified via
  Eq.~(\ref{coupledbrane}) by inserting the ansatz for our metric, and choosing
  a static gauge for the coordinates parametrising the world-volume  
  \cite{DufSte:91,DufKhuLu:94}.
  We see that the Lagrangian gives a potential, which due to the chosen 
  linear interdependence of the functions $A,B, and C,$
  vanishes\footnote{Recall that this results from  
  supersymmetry arguments.  We should note that the no-force condition is
  not a necessary condition for the reduction, since we are constructing
  an infinite array.  Hence, though unstable, the gravitational and
  electrical forces on each individual $p$-brane cancel. We can
  therefore also construct multi-centre solutions for
  non-supersymmetric cases \cite{LuPopSte:96}}. 
  Mathematically this amounts to the validity of 
  superimposing solutions to Laplace's equation (\ref{laplace}).  
  
  The above formulae for a single M2-brane lead to 
  \cite{DufSte:91}:
\begin{eqnarray} \label{M2brane}
  ds^2&=&H(r)^{-\frac{2}{3}}dx^{\mu}dx^{\nu}\eta_{\mu\nu}
       +H(r)^{\frac{1}{3}}dy^m dy^m \nonumber \\
  C_{012}&=&H(r)^{-1} \nonumber \\
  H(r)&=&1-\frac{k_3}{r^6}.
\end{eqnarray}

  Choosing ${\bf \hat{y} }_8={\bf \hat{z}}$ as the transverse axis of 
  compactification, the superposition of solutions gives
\begin{equation} \label{plainsum}
H({\bf y}) =  1+k_3 \sum_{n \in {\bf Z} } \frac{1}{\left| {\bf y} - n a
    {\bf \hat{z}} \right|^6},
\end{equation}
  a set of parallel membranes of the same orientation, with the same mass
  and charge; they are located periodically along the ${\bf \hat{z}}$ axis
  with period $a=2\pi R_{11}$.  Identifying the membranes we change 
  the topology, ${\bf R}^{11} \rightarrow {\bf R}^{10} \times S^1$.
  
  Performing a change of variables:
\begin{eqnarray} \label{Hmulti}
  \left| {\bf y} - n a {\bf \hat{z}} \right|^6 &=& \left(\underbrace{
  (y^1)^2 + \ldots +(y^7)^2}_{\hat r^2} + (\underbrace{y^{8}}_{z}-n
  a)^2 \right)^3 \nonumber \\
  H(\hat{r},z)&=&1+k_3 \sum_{n \in {\bf Z} } \frac{1}{\left(\hat r^2 +
           \left( z - n a \right)^2 \right)^3}.
\end{eqnarray}
  The explicit dependence on the compactified coordinate $z$ poses a problem
  for the standard Kaluza-Klein procedure.  This can be ameliorated by
  letting $R_{11} \rightarrow 0$, creating a smeared out brane along
  ${\bf \hat{z}}$ and replacing the sum by an integral:
\begin{equation}
  \sum_{\alpha}\frac{k_3}{\left|{\bf y}-{\bf y}_\alpha \right|^6}
  \longrightarrow \int_{-\infty}^{\infty}\frac{k_3 dz}{(\hat
  r^2+z^2)^3} = \frac{\tilde k_3}{\hat r^5},
\end{equation}
  where $\tilde{k_3}=\frac{3\pi k_3}{8 r^5}$
  \cite{LuPopSte:96}, giving  
\begin{equation} 
  ds^2=\left(1+\frac{\tilde k_3}{\hat r^5}\right)^{-2/3}dx^{\mu}dx^{\nu}\eta_{\mu\nu}
  +\left(1+\frac{\tilde k_3}{\hat r^5}\right)^{1/3}(dz^2+dy^{\tilde{m}}dy^{\tilde{m}}).
\end{equation}

  Using Eq.~ (\ref{reductionmetric}), and the values for $\alpha$ and $\beta$
  we have
\begin{equation} \label{KKansatz}
   ds^2_{11}=e^{-\frac{\varphi}{6}}ds^2_{10}+e^{\frac{4\varphi}{3}}dz^2,
\end{equation}
  giving
\begin{eqnarray} \label{D2brane}
  ds^2_{10}&=&\left(1+\frac{\tilde{k}_3}{\hat r^5}\right)^{-\frac{5}{8}}
                   dx^{\mu}dx^{\nu}\eta_{\mu\nu}
              \left(1+\frac{\tilde{k}_3}{\hat r^5}\right)^{\frac{3}{8}}
                   dy^m dy^m \nonumber \\
  e^{\phi}&=&\left(1+\frac{\tilde{k}_3}{\hat r^5}\right)^{1/4},
\end{eqnarray}
  where the Kaluza-Klein dilaton $\varphi$ becomes the IIA dilaton field
  $\phi$.  The solution matches with the result of
  the Type IIA $D=10$ supergravity equations (\ref{pbranesoln})
  \cite{LuPopSezSte:95,DufLu:92}.

  The smearing of the brane is effectively equal to 
  ignoring all the heavy modes along the ${\bf \hat{z}}$ direction, which die
  off exponentially fast\cite{LuPop:97b}. Hence at small compactification 
  radii, or equivalently $r \gg R_{11}$, we can see only the above
  solution. This solution turns out to be singular at $r=0$, 
  as can be seen by calculating the Ricci scalar at that point (see 
  below).  We expect that including the heavy modes will remove
  the singularity.  This amounts to finding the explicit $z$
  dependence of the prefactor: turning to Eq.~(\ref{Hmulti}),  
  writing $H(\hat r,z)=1+k_3 g(\hat r,z)$, with 
  $g(\hat r,z)=-\frac{1}{4\hat r}\frac{\partial}{\partial \hat r}
     \left(-\frac{1}{2\hat r}\frac{\partial f}{\partial \hat r}\right)$
  and $f(\hat r,z)= \sum_{n \in {\bf Z} } \frac{1}{\left(\hat r^2 +
               \left( z - n a \right)^2 \right)^3} $, we can
  evaluate $H(\hat r,z)$ by evaluating $f$.  This has been done 
  explicitly in \cite{HarShe:78,Myers:87}:
\footnote{ Letting $f(z)=\phi(z)\pi\cot(\pi z)$ we can use simple
 complex analysis to evaluate the sum:
\begin{equation}
  \oint \limits_{\gamma=\lim \limits_{N\rightarrow\infty}\gamma_N} f(z) =
  2\pi i\left(\sum_{n=-\infty}^{\infty}\phi(n)-
  \mathrm{Res}_{z\not\in{\bf Z}},f(z)\right),
\end{equation}
  where $\gamma_n$ is the square passing through
  $(\pm\left(N+\frac{1}{2}\right),0)$ and
  $(0,\pm i\left(N+\frac{1}{2}\right))$
  \cite{CarKroPea:83}.
}
\begin{equation}
  f(\hat r,z)=\frac{1}{2 R_{11} \hat r}\frac{\sinh(\hat r/R_{11})}{\cosh(\hat r/R_{11})- 
         \cos(z/R_{11})}.
\end{equation}
  Hence
\begin{eqnarray} \label{compactsoln}
  H(\hat r,z) & = & 1+\frac{3 k_3}{16 R_{11}\hat r^5}\frac{\sinh
  \hat r/R_{11}}{\cosh \hat r/R_{11} - \cos z/R_{11}} \nonumber \\
  &&- \frac{3 k_3 }{16 R_{11}^2 \hat r^4}
  \frac{1-\cosh \hat r/R_{11} \cos z/R_{11}}{\left(\cosh \hat r/R_{11}-
  \cos z/R_{11} \right)^2} \nonumber \\
  & & -\frac{k_3}{16 R_{11}^3 \hat r^3}\frac{\sinh \hat r/R_{11} \left(2-\cos^2
  z/R_{11}-\cosh \hat r/R_{11} \cos z/R_{11} \right)}
  {\left(\cosh \hat r/R_{11} - \cos z/R_{11} \right)^3}. \nonumber \\
\end{eqnarray}
 
  It is this expression that will be helpful in realizing what the 
  eleven-dimensional character of the D2 brane is, in comparison with the
  picture one has of the fundamental string, a tube\cite{DufHowInaSte:87}.

  First, however, we will compare the limiting expression of 
  Eq.~(\ref{compactsoln}) to see how it appears in ten dimensions,
  before we look at the eleven dimensional effects.

\section{Ten Dimensional Solutions} \label{sec:10dptofv}
 We have three starting points for the ten dimensional picture of the 
 D2-brane.  First we can solve the Type IIA supergravity equations 
 explicitly, giving solutions of the form (\ref{pbranesoln}).  The next
 approach consisted in smearing out the brane and we reobtained the same 
 expression for the D2-brane.  We are left with the third intrinsically
 eleven-dimensional expression (\ref{compactsoln}), which has to be
 analysed in the appropriate limiting situations.
 
 The limits of interest are the two extremal cases of letting the 
 compactification radius $R_{11} \rightarrow 0$; and approaching
 the brane up close in which case $\hat r,z$ become of O($R_{11}$).
 We expect the former limit to yield the prefactor of
 Eq.(\ref{D2brane}), reproducing the ten dimensional solution,
 barring any conformal scaling.
 
 As $R_{11} \rightarrow 0$, the exponential parts of (\ref{compactsoln}) 
 conspire to leave only the leading order term
\begin{equation} \label{Limprefac}
   H(r)_{\mathrm{multi}}=1+\frac{3k_3}{16R_{11}\hat r^5}.
\end{equation}
 Using this asymptotic form of the prefactor and applying the standard
 Kaluza-Klein ansatz (\ref{KKansatz}), we obtain the same form for the
 D2-brane, as in the previous two cases.  
 The difference in these three solutions stems from the constant of 
 integration, which represents the charge and mass of the membranes 
 considered.  
 
 Using Eqns.~(\ref{D2brane}) for the D2-brane, 
 we can calculate the ADM mass and charge explictly
 for the three cases mentioned (\ref{ADMmass}):
\begin{equation} \label{BPS}
 \mathcal{M}=\frac{5}{4} \tilde{k},\quad \mathcal{Q}=\frac{5}{4} \tilde{k},
\end{equation}
 saturating the Bogomol'nyi inequality, indicating the
 supersymmetry of the solution.

 Inserting the three different cases into these expressions, we
 find how the reduction affects our interpretation of the situation.
 In the solution coming directly from the IIA equations, $k_3^{(10)}$ is
 completely arbitrary
 \footnote{The superscripts will indicate the spacetime dimensions from
 which these constants arise.}.  The continuum limit gave 
 $\tilde{k}_3^{(10)}=\frac{3\pi k_3^{(11)}}{8}$, where the multiplicative factor
 has no physical importance since it can be absorbed into the
 eleven-dimensional constant of
 integration.  Both these cases give us a continuous spectrum of masses
 for the D2-brane.  We have to invoke the Dirac quantisation, using
 D2 and NS5 brane ``electric-magnetic'' duality, to give us discrete 
 charges and by (\ref{BPS}), discretising the BPS-spectrum. 
 
 However, the third reduction takes the limit $R_{11} \rightarrow 0$ 
 after the sum has been evaluated
 explicitly.  In this fashion we obtain an $\frac{1}{R_{11}}$
 dependence in the ADM mass.  Though we still have to apply Dirac quantisation
 to make $\tilde{k}_3 \sim \mathcal{M} \sim n, n\in {\bf Z}$, this
 $\frac{1}{R_{11}}$ dependence implies a Kaluza-Klein tower of BPS states, 
 which would be too pathological if continuous.  Further this dependence 
 gives the correct D-brane coupling to the dilaton, $e^{\phi}$, unlike
 the usual Kaluza-Klein approach.  
 This can be seen by following Witten's argument \cite{Witten:95}, where
 we use conformal scaling and Kaluza-Klein reduction to relate the radius
 of the compactified dimension in the two different metrics used to 
 measure it; the eleven dimensional metric, and the string metric.
 

 We may now ask what the reverse procedure ``Kaluza-Klein oxidation''
 produces when applied to Eq.~(\ref{D2brane}).  Since this solution
 starts from a continuum distribution of branes, we may expect that
 oxidation brings us back to it.  However this need not be so. The most 
 general resulting harmonic prefactor
 in the eleven-dimensional metric due to such oxidation is of the
 form:
\begin{eqnarray}
 H(\hat r,z) &=& \sum_m f_m(\hat r) e^{imz/R_{11}} \nonumber \\
 f_m&=&\frac{c_m}{\hat r^{\tilde{d}/2}}
     K_{\tilde{d}/2}\left(\frac{|m|\hat r}{R_{11}}\right)+
     \frac{\tilde{c}_m}{\hat r^{\tilde{d}/2}}
     I_{\tilde{d}/2}\left(\frac{|m|\hat r}{R_{11}}\right),
\end{eqnarray}
 where $c_m$ and $\tilde c_m$ are arbitrary.  The heavy modes, $m\neq 0$, 
 exponentially die of as 
 $\hat r\rightarrow\infty$ and $f_0$ is the original ten dimensional
 prefactor\cite{LuPop:97b}.  There is a slight ambiguity in what
 the expansion should be (if a periodic array at all), as the imprint at
 $r=0$ is ambiguous \cite{Myers:87}.  
 The above discussion favours the periodic array.
 
 Moving to the other limiting case of 
 $\hat r,z \sim \mathrm{O(}R_{11}\mathrm{)}$,the non-zero modes should
 effectively boost the apparent ten dimensional
 solution to eleven dimensions.  This limit, for a small compactification
 radius is equivalent to letting $\hat r$ become small.  Ignoring the heavy modes
 Eq.~(\ref{D2brane}) gives a Ricci scalar that blows up in the limit $\hat r\rightarrow 0$:
\begin{equation}
  \lim_{\hat r\to 0}R_{D2} = \lim_{r\to 0}\frac{75}{32}
  \left(1+\frac{\tilde{k}_3}{\hat r^5}\right)^{-\frac{19}{8}}\hat r^{-12}\tilde{k}_3^2
  \rightarrow \infty,
\end{equation}
 identifying a naked singular membrane sheet at the origin.  It may be
 interpreted as a source to the field equations.
 
 On the other hand, a similar calculation for the M2-brane 
 (\ref{M2brane}) leads to the eleven-dimensional result
\begin{equation} \label{RicciM2}
  \lim_{r\to 0}R_{M2} = \lim_{r\to 0}6 k_3^2 
  \left(r^6+k_3\right)^{-\frac{7}{3}}=6k_3^{-\frac{1}{3}},
\end{equation}
 hinting that $r=0$ for the M2-brane might be a horizon 
 \cite{GibTow:93,DufGibTow:94}, making the non-zero modes
 a relevant contribution to smearing out the singularity in the D2-brane.
 This also justifies the quantisation of such a membrane object in accordance
 with \cite{Townsend:95a}, as it would counter the ability of the
 membrane to deform at zero energy cost.  One might obtain a discrete
 spectrum for both the M2 and D2-brane.

 Another point of view to the singularity of the D2-brane
 (\ref{D2brane}) is its dependence on the correct choice of metric
 \cite{DufLu:94,DufKhuLu:92}.  Eleven-dimensional supergravity does
 not have a dilaton introducing ambiguities in the
 metric due to conformal rescaling.  In ten dimensional string theory,
 however, we have the frame of the object considered (e.g.~the D2-brane);
 the Einstein frame that removes the dilatonic prefactor of the Ricci 
 term in the Lagrangian; and the dual frame, which results from performing
 an ``electric-magnetic'' duality transformation
 \cite{DufKhuLu:92,DufGibTow:94}.  All these frames
 are related by a conformal rescaling of the metric.  It is these rescalings 
 that remove the singularities in the D2-metrics, as can be checked by 
 calculating the Ricci scalar and the proper time it takes for fundamental 
 (dual) objects to fall into their dual counterpart.
 The proper time is infinite in such cases, indicating that the 
 singularity is physically irrelevant.  We are therefore always in the 
 position of making the singularity vanish in accordance with
 the eleven-dimensional singularity free horizon at $r=0$.

\section{Local and Global Structure}
 We may now take the opposite limit from
 the previous section, i.e.~approaching the D2-brane up close with
 both $\hat r,z\rightarrow 0$. In doing this 
 we must be careful in which order we let the coordinate variables approach
 zero.  There are three cases to be discerned: $\hat r\rightarrow 0, z=0$;
 $\hat r=0,z\rightarrow 0$; and a diagonal incoming trajectory
 $\hat r=\lambda z,z\rightarrow 0$.  The second case might seem dubious from
 Eq.~(\ref{compactsoln}), however, it is easy to see from (\ref{Hmulti}) that
 $\hat r=0$ is allowed.
 In all three situations the form of the metric
 prefactor is of the form
\begin{equation}
 H_{\mathrm{multil}}\rightarrow 1+\frac{k_3}{\mathrm{R}^6},
\end{equation}
 where R is the distance from $(\hat r,z)=(0,0)$.  Up close we 
 reobtain the M2-brane. 
 
 To show that Eq.~(\ref{RicciM2}) indeed reflects the non-singularity of
 a single M2-brane at $r=0$, we can construct an analytic coordinate 
 extension past this horizon for Eqns.~(\ref{M2brane}), 
 following Refs \cite{GibTow:93,DufGibTow:94}: 
\begin{eqnarray}
 r&=&k_3^{1/6}(\Phi^{-3}-1)^{-1/6} \label{coordtrans} \\
 ds^2&=&\left[\Phi^2(-dt^2+{dx^1}^2+{dx^2}^2)+\frac{k_3^{1/3}}{4}\Phi^{-2}d\Phi^2+k_3^{1/3}d\Omega_7^2\right] \label{line1} \\
 &&+\frac{k_3^{1/3}}{4}\Phi^{-2}\left[(1-\Phi^3)^{-7/3}-1\right]d\Phi^2 \nonumber \\
 &&+k_3^{1/2}\left[(1-\Phi^3)^{-1/3}-1\right]d\Omega_7^2 \nonumber \\
 A_{\mu\nu\rho}&=&\Phi^3\epsilon_{\mu\nu\rho},
\end{eqnarray}
 where 
\begin{center}
\begin{tabular}{ll}\
   $r\rightarrow \infty$ & $\Phi \rightarrow 1^{-}$ \\
   $r=0$, the horizon    & $\Phi=0$              \\
   into the horizon      & $\Phi<0$              \\
\end{tabular} \\ 
\end{center} 

 To see that we can pass through the horizon in these coordinates, we
 look at the near horizon geometry, i.e.~$\Phi\approx 0$, which is
 AdS$_4 \times$S${}^7$, line (\ref{line1}).  We can see this by taking
 AdS$_4$ as the quadric in ${\bf R}^{3,2}$ given by
\begin{eqnarray} \label{AdSmet}
  (X^0)^2+(X^4)^2-(X^1)^2-(X^2)^2-(X^3)^2=\frac{k_3^{1/3}}{4} \nonumber\\
  ds^2_{\mathrm{AdS}_4}=-d(X^0)^2-d(X^4)^2+d(X^1)^2+d(X^2)^2+d(X^3)^2,
\end{eqnarray}
 with
\begin{eqnarray}
 (X^4-X^3)&=&\Phi \nonumber \\
 X^0&=&t\Phi      \nonumber \\
 X^1&=&x^1\Phi \nonumber \\
 X^2&=&x^2\Phi \nonumber \\
 (X^4+X^3)&=&\frac{k_3^{1/3}}{4}\Phi^{-1}+({x^1}^2+{x^2}^2-t^2)\Phi,
\end{eqnarray}
 which reproduces (\ref{line1}) after insertion into (\ref{AdSmet}).  $\Phi$ is an
 analytic function on AdS${}_4$ and can be continued through to negative
 values.  Though it seems that (\ref{line1}) goes bad, it is the coordinates
 $\{t,x^1,x^2\}$ that fail.  The metric can be brought to a healthy form
 at $\Phi=0$ by changing the metric to the coordinates 
 $\{X^0,X^1,X^2,X^3\}$. The higher order terms also depend
 analytically on $\Phi$ and can also be continued through to negative values of $\Phi$.
 $\Phi=0$ is a horizon.  Indeed there is a coordinate 
 singularity at $\Phi=-\infty$. Since $r=0, \Phi=0$ is made up of two 
 connected components \cite{DufGibTow:94}, we have two horizons, and 
 we can continue through them separately, obtaining a Carter-Penrose 
 diagram similar to the Reissner-Nordstr\"{o}m extreme black-hole 
 \cite{GibHorTow:94}.

 Before going on, we first would like to introduce isotropic coordinates
 in the interior of a single M2-brane.  The coordinate transformation (\ref{coordtrans}) 
\begin{equation}
 r=k_3^{1/6}\Phi^{1/2}(1-\Phi^3)^{-1/6}
\end{equation}
 at first sight seems to produce a non-analytic extension because of
 the fractional power of $\Phi$.  However, only even powers of $r$ appear
 throughout the metric (\ref{M2brane}), eliminating this problem.  We
 also do not have to worry about the complexification of $r$ as 
 $\Phi$ passes through the horizon, becoming negative.  For this range
 define $\Phi=-\Phi_1<0$, such that 
 $r^2=-r_1^2=-k_3^{1/6}\Phi_1^{1/2}(1+\Phi_1^3)^{-1/6}<0$. By changing 
 to these coordinates the M2-brane metric becomes
\begin{equation}
 ds^2_{r^2<0}=(\frac{k_3}{r_1^6}-1)^{-2/3}dx^{\mu}dx^{\nu}\eta_{\mu\nu}+
            (\frac{k_3}{r_1^6}-1)^{1/3}(dr_1^2+r_1^2d\Omega_7^2)
\end{equation}
 for $r_1>0$.  Continuing past $r=0=r_1$, we encounter a curvature
 singularity at $r_1=k3^{1/6}$, i.e.~at $\Phi=-\infty=-\Phi_1$.  This can
 be inferred from the blow up of the Ricci curvature scalar, which becomes
 infinite there.
 
 Expanding on the above analogy with $D=4$ extremal black-holes, we can 
 place several extreme Reissner-Nordstr\"{o}m black holes in space time
 \cite{Majumdar:47,Papapetrou:47,HarHaw:72}, resulting in a static
 configuration.  This is what our higher-dimensional counterpart of 
 superimposed M2-branes is.  From \cite{HarHaw:72} we know how the 
 resulting space-time should look and a similar construction for
 black holes in $N+1$ dimensions has been performed in \cite{Myers:87}.
 We can now proceed along the lines of \cite{GibHorTow:94} to see how
 the non-zero modes due to the presence of the other black holes affects
 the near-horizon geometry.  
 
 Following \cite{GibHorTow:94} we take Eq.~(\ref{plainsum}) and concentrate
 on the brane located at $n=0$
 \footnote{
   $n=0$ can be chosen to label any arbitary M2-brane, as the stacking
 is infinite.
}.  Instead of using the $(\hat r,z)$ split, we 
 take the eleven dimensional isotropic radial coordinate and rewrite
 the sum as:
\begin{equation} \label{LegendreSum}
   H=1+\frac{k_3}{r^6}+k_3\sum_{n\neq 0}\frac{1}{(r^2+(an)^2-2anr\cos\theta)^3},
\end{equation}
 where $\theta$ is the polar angle along the ${\bf y}_8$-axis.
 We can therefore rewrite the sum in terms of $C^{3}{}_l(\cos\theta)$,
 ultraspherical polynomials, which form a complete 
 set of harmonics on S${}^{d-2}$:\footnote{
 The ultraspherical polynomials are given by the generating function
 \begin{equation}
   \frac{1}{(1-2tx-t^2)^{\alpha}}=\sum_{n=0}^{\infty}C^{\alpha}{}_l(x)t^n.
 \end{equation}
 $\alpha=1/2$ gives the Legendre Polynomials, for $\alpha=0,\
 1$ we obtain the Tshebyscheff polynomials.  
 In our case $\alpha=3$. \cite{Arfken:85}
}
\begin{equation}
 \sum_{l=0}^{\infty}a_{l}r^{l}C^{3}{}_l(\cos\theta),
\end{equation}
 where the expansion requires $r<an$.  Since we are concentrating on
 the M2-brane located at $n=0$, this condition is satisfied.
 To continue through $r=0$ we introduce the same variable as before:
\begin{eqnarray}
 r&=&k^{1/6}\Phi^{1/2}(1-\Phi^3)^{-1/6} \nonumber \\
  &=&f(\Phi)\Phi^{1/2},                  \label{rtransform}
\end{eqnarray}
 where $f(\Phi)$ is an analytic function of $\Phi$ at $\Phi=0$.  Hence
 the prefactor becomes
\begin{eqnarray}
 H&=&\Phi^{-3}+\sum_{l}^{\infty}a_{l}(\Phi)\Phi^{l/2}C^3_{l}(\cos\theta) \nonumber \\
  &=&\Phi^{-3}\left[1+\sum_{l=6}^{\infty}b_{l}(\Phi,\Omega)\Phi^{1/2}\right],
\end{eqnarray}
 where $a_{l}$ are analytic functions of $\Phi$ and the 
 $b_{l}$ are analytic functions of $\Phi$ and of $S^7$.  The
 leading term reproduces the above discussion of 
 AdS${}_4\times$S${}^7$:  
\begin{eqnarray} \label{LeftOver}
 ds^2&\stackrel{\Phi\rightarrow 0}{\sim}&\Phi^2(-dt^2+d{x^1}^2+d{x^2}^2)+
   k_3^{1/3}\Phi^{-2}d\Phi^2+k_3^{1/3}d\Omega_7^2 \nonumber\\
 &&+f(\Phi)(-dt^2+d{x^1}^2+d{x^2}^2)+g(\Phi)d\Phi^2+h(\Phi)d\Omega_7^2,
\end{eqnarray}
 where $f(\Phi),g(\Phi),h(\Phi)\stackrel{\Phi\sim 0}{\rightarrow}0$, as
 they include terms of O($\Phi$). 
\footnote{
 We see that $(r,z)=(0,0)$ is not a singularity, and one can also calculate
 the Ricci scalar at that membrane surface to give $R=\frac{12}{k^{1/3}}$
 indicating the possibility of passing through the horizon.
}
 The higher order terms, however, include 
 powers of $\Phi^{1/2}$, similarly to the single M2-brane, 
 which seem to prevent an analytic continuation through
 $r=0=\Phi$ \cite{GibHorTow:94}.  Because of this apparent lack of
 analyticity, it seems there exists no unique extension across the
 horizon. One can match onto essentially any solution of the form
 (\ref{M2brane}) and (\ref{plainsum}) with the same total mass.  Such 
 a situation also arises in the case of dynamical multi-black holes 
 in five dimensional de Sitter spacetime \cite{BriHorKasTra:94} and  
 higher dimensional multi-$p$-branes in a static spacetime 
 \cite{GibHorTow:94}.
 The smoothness of such solutions in a static environment was analysed 
 in \cite{Welch:95} removing the possibility of interpreting the finite 
 differentiability of the metric as a result of gravitational radiation.
 Whether or not this lack of smoothness has a physical meaning, 
 as the metric is always 
 suitably differentiable in these cases ($C^k,k\ge 2$), was discussed 
 for an exact solution
 in Ref.~\cite{CruSin:92}.  There it was argued that an observer
 could in principle keep track of the derivatives of the Riemann tensor,
 hence detecting when he has crossed the horizon.

 However, from the physical point of view, we know that every single static
 black hole has a smooth analytic horizon.  We would like this to hold
 for its dimensionally reduced counterparts and for spacetime
 backgrounds other 
 than flat Minkowski spacetime.  To see how this affects
 our current solution we may expand our series (\ref{LegendreSum}) in 
 terms of the above ultraspherical polynomials.  The coefficients $a_l$
 become:
\begin{equation}
  a_l=\sum_{n\neq 0}\frac{1}{(an)^{l+6}}\quad\left\{\begin{array}{l} =0,\ l\
  \mathrm{odd}\\ =\frac{2\zeta(l+6)}{a^{l+6}},\ l\ \mathrm{even} \end{array}\right. ,
\end{equation}
 where $\zeta(s)$ is the Riemann Zeta function.
 After inserting (\ref{rtransform})
 into the expansion, we see that the power series contains only integral
 powers of $\Phi$.  The resulting $f(\Phi),g(\Phi),h(\Phi)$ in 
 Eq.~(\ref{LeftOver}) can be analytically
 continued through to negative values of $\Phi$, hence negative values of
 $r^2$.  This analytic continuation should not surprise us, as it 
 was speculated in Ref.~\cite{BriHorKasTra:94} that for multi-black hole solutions 
 the differentiability of the horizon is increased if the masses are so
 arranged that the first $n$ multiple moments vanish, which for our
 case of infinite ``extremal black-holes'' would certainly be the case.
 This has been verified for a low number of black-holes in Ref.~
 \cite{Welch:95}.
 
 As in the discussion for the single M2-brane, the power series of the 
 prefactor contains only even powers of $r$, so that we can proceed to 
 formally make the change of variables $(\hat r,z)\rightarrow(i\hat r,iz)$.
\begin{eqnarray} \label{M2inside}
  H(\hat r,z) & = & 1+\frac{3 k_3}{16 R_{11}\hat r^5}\frac{\sin
  \hat r/R_{11}}{\cos \hat r/R_{11} - \cosh z/R_{11}} \nonumber \\
  &&- \frac{3 k_3 }{16 R_{11}^2 \hat r^4}
  \frac{1-\cos \hat r/R_{11} \cosh z/R_{11}}{\left(\cos \hat r/R_{11}-
  \cosh z/R_{11} \right)^2} \nonumber \\
  & & +\frac{k_3}{16 R_{11}^3 \hat r^3}\frac{\sin \hat r/R_{11} \left(2-\cosh^2
  z/R_{11}-\cos \hat r/R_{11} \cosh z/R_{11} \right)}
  {\left(\cos \hat r/R_{11} - \cosh z/R_{11} \right)^3}.\nonumber \\
\end{eqnarray}
 The metric becomes:
\begin{equation}
  ds^2=(-H(\hat r,z))^{-\frac{2}{3}}dx^{\mu}dx^{\nu}\eta_{\mu\nu}
       +(-H(\hat r,z))^{\frac{1}{3}}(d\hat r^2+\hat r^2d\Omega_6^2+dz^2).
\end{equation}
 
 Here we have both $\hat r,z>0$, though the origin does not correspond to a point.
 We can see this by looking at the $d\Omega_7^2$ 
 coefficient, which is non-zero.  This is to be expected, as $(\hat r,z)=(0,0)$
 is the membrane horizon through which we analytically extended.  
 We can now proceed inward with $(\hat r,z)$ to find the singularity  
 that was 
 previously located at $\Phi=-\infty$.  The location of the
 singularity is at $H(\hat r,z)=0$, which indeed has zeroes for
 suitable values of $(\hat r,z)$.  To see that at this location we have a true
 singularity, we can follow the the field invariant 
 $J=G^2=\left(\frac{\nabla H}{H^2}\right)^2$, which diverges for $H=0$.
 
 Having found the hidden singularity beyond the horizon we look at the
 structure of the space that the external observer cannot see, $\Phi<0$.
 We note that the $z$ coordinate loses its periodicity
 when entering the horizon.  However, we are not surprised by this,
 since the periodicity existed outside the membrane horizon separating
 ``inside'' from ``outside.''  Having the periodicity removed also makes
 the limit of $R_{11}\rightarrow 0$ become irrelevant, as there is no radius.
 This seems to be in accordance with the ``outside'' limit becoming singular at the
 horizon in the lower-dimensional limit, disconnecting the ``inside''
 from any physical
 observer who lives at a safe distance from the naked singularity, hence
 removing the physical relevance of the ``inside'' region.  However,
 as stated above, this singularity is removable, depending on which
 metric we choose to use in ten dimensions.  Since our vertical dimensional
 reduction keeps the membrane object as the ``fundamental'' element in the
 lower dimensional theory, this singularity seems to be quite real from
 the perturbative expansion point of view of the IIA theory in this 
 particular vacuum\cite{DufKhuLu:92}.

 Another way to interpret the limit of $R_{11} \rightarrow 0$, taken
 outside the horizon, is to 
 say that $\hat r\gg R_{11}$.  We see the eleventh dimension 
 only becoming 
 relevant close to the membrane.  From Eq.~(\ref{M2inside}) this dimension
 becomes irremovable inside and we are drawn to the speculation that
 this confinement of the eleventh dimension, might be related to the 
 confinement of the gauge field of a D2-brane to its worldvolume.  This
 would be in accordance with the duality transformations performed in
 Ref.~ \cite{Townsend:95b}, where the eleventh dimension becomes 
 the gauge field living on the world-volume.

\pagebreak

\end{document}